\documentclass[%
reprint,superscriptaddress,
amsmath,amssymb,
pre,
aps, 
showkeys,
]{revtex4-2}

\usepackage{hyperref}
\usepackage{graphicx}
\usepackage{amsfonts} 
\usepackage{thmbox} 
\usepackage{empheq} 
\usepackage{version}
\usepackage{times}
\usepackage{makeidx}
\usepackage{enumerate}
\usepackage{amsmath}
\usepackage{fancyhdr}
\usepackage{color}

\newcommand{\beq}{\begin{equation}}     \newcommand{\eeq}{\end{equation}}
\newcommand{\beqa}{\begin{eqnarray}}    \newcommand{\eeqa}{\end{eqnarray}}
\newcommand{\bde}{\begin{description}}  \newcommand{\ede}{\end{description}}
\newcommand{\ben}{\begin{enumerate}}    \newcommand{\een}{\end{enumerate}}

\newcommand{\kT}{{k_{\rm B}T} } 
 
\newcommand{\eqn}[1]{\beq{ #1 }\eeq}

\newcommand{\inv}[1]{{\frac{1}{#1}}}

\newcommand{\inRbracket}[1]{{\left({#1}\right)}}

\newtheorem[L]{thm}{Theorem}[section]
\newtheorem{cor}[thm]{Corollary}
\newtheorem{theorem}{{\sf Assertion :}}[section] 
\newtheorem{definition}{\sf Definition} 
 
\newtheorem{lemma}[theorem]{Lemma}
\newcommand{\bth}{\begin{thm}}  
\newcommand{\blem}{\begin{lemma}}  
\newcommand{\elem}{\end{lemma}}  
\newcommand{\bpr}{\begin{proof}}  
\newcommand{\epr}{\end{proof}} 
\newcommand{\bdefine}{\begin{definition}}  
\newcommand{\edefine}{\end{definition}}  
\newcommand{\bcor}{\begin{cor}} 
\newcommand{\ecor}{\end{cor}}  
\newcommand{\bprop}{\begin{example}[Property]}  
\newcommand{\eprop}{\end{example}}  
\newcounter{formulaire}
\newcommand{\beqf}{\addtocounter{formulaire}{1}\begin{equation}}
\newcommand{\eeqf}{\tag{R \arabic{formulaire}}\end{equation}}
\newcommand{\beqaf}{\addtocounter{formulaire}{1}\begin{equation}\begin{array}{rcl}}
\newcommand{\eeqaf}{\end{array}\tag{R \arabic{formulaire}}\end{equation}}

\newcommand{\meq}{{m^{(\rm eq)}}}
\newcommand{\Ppq}{{P^{\rm  (PQ)}}}
\newcommand{\Peq}{{P^{\rm  (eq)}}}

\usepackage{dcolumn}

\usepackage{bm}

\usepackage{listings} 

\begin{document}
	
\title{Martingale-induced local invariance in progressive quenching}

\author{Charles Moslonka}
\affiliation{Laboratoire Gulliver, UMR CNRS 7083, ESPCI Paris, Université PSL\\
	10 rue Vauquelin, 75005, Paris, France.}
\email[Corresponding author: ]{charles.moslonka@espci.psl.eu}
\author{Ken Sekimoto}
\affiliation{Laboratoire Gulliver, UMR CNRS 7083, ESPCI Paris, Université PSL\\
	10 rue Vauquelin, 75005, Paris, France.}
\affiliation{Laboratoire Matière et Systèmes Complexes, UMR CNRS 7057, Université de Paris,\\
	10 Rue Alice Domon et Léonie Duquet, 75013, Paris, France }
\email[Corresponding author: ]{ken.sekimoto@espci.psl.eu}

\date{\today}
	
\begin{abstract}
	
	Progressive quenching (PQ) is a stochastic process during which one fixes, one after another, the degrees of freedom of a globally coupled Ising spin system while letting it thermalize through a heat bath. 
	It has previously been shown that during PQ, the mean equilibrium spin value follows a martingale process and this process can characterize the memory of the system. 
	In the present study, we find that the aforementioned martingale implies a local invariance of the path weight for the total quenched magnetization, the Markovian process whose increment is the spin that is fixed last. 
	Consequently, PQ lets the probability distribution for the total quenched magnetization evolve while keeping the Boltzmann-like factor, or a canonical structure, under constraint, which consists of a path-independent potential and a path-counting entropy. 
	Moreover, when the PQ starts from full equilibrium, the probability distribution at each stage of PQ is found to be the limit distribution of what we call recycled quenching, the process in which a randomly chosen quenched spin is unquenched after a single step of PQ. The local invariance is a consequence of the martingale property, and not an application of known theorems for the martingale process.
	
\end{abstract}

\maketitle
\section{Introduction}

Martingales \cite{martingale-book} have been widely known to physicists as a useful tool for studying stochastic processes. By converting  a stochastic process into a martingale, one can use many theorems derived from probability theory, which allow to reach those results that would otherwise be difficult or laborious to obtain.
Recently in non-equilibrium statistical physics, it was recognized that
some variables of physical significance can be reinterpreted in light of martingales. 
In such cases the martingale property brings directly consequences of physical interest.

The first and now widely known case is the path probability ratios appearing in a variety of fluctuation theorems or non-equilibrium equalities. 
The authors of \cite{martingale-Gupta2011,Roldan-prX2017} brought to the physicists' attention  that such ratios are recognized by mathematicians as the Radon-Nikodym derivative and that they are martingale processes. Their work improved
the understanding of entropy production as an action functional 
and allowed to introduce the concept of stopping time \cite{Edgar-stopT,Edgar-Pekoa-stopping-time-PRL2019,Edgar-QM-stopping-time-PRL2019,Neri-stopT}, such as the random cycle duration of autonomous mesoscopic heat engine \cite{sar}.

The second case is what we call Progressive Quenching (PQ). In this process a globally interacting spin system (the Curie-Weiss model) undergoes the fixation - or \textit{quenching} - of an Ising spin one after another with a sufficient time interval so that the unquenched spins remain in equilibrium with a heat bath. This progressive cool-down of a systems' degrees of freedom might be seen as a simplification of certain processes. For example when a molten material is pulled out from a furnace and is quickly cooled down, the fluid degrees
of freedom associated to fluid particles are progressively fixed
(quenched)\cite{Damien2017}. On a more socially-oriented point of view, we might also
consider the process of decision making by a community in
which each member progressively makes up their mind
before a referendum. In both examples, the already fixed
part can influence the behavior of the part whose degrees of
freedom are not yet fixed. Moreover, this paradigm could also be applied to study the evolution of mechanical properties of certain materials because the elasticity is long-ranged. For example ripples propagates in graphene sheets \cite{meyer2007structure} with quenched defects. A model of spins interacting indirectly through an elastic string has been studied in \cite{bonilla2012ripples}.

We found that the evolution of the \textit{mean equilibrium spin of the unquenched part} constitutes a martingale process \cite{PQ-KS-BV-pre2018}, where the discrete time is represented by the number of quenched spins.
While the context here is more specific than in the first case, the mechanism leading to the martingale property is not the Radon-Nikodym derivative but the ``tower-rule'' or law of total expectation  (see below).
In this system of PQ the martingale property leads to a persistent memory by which 
we could infer the past data \cite{PQ-KS-BV-pre2018} or predict approximately the future distribution \cite{PQ-CM-KS-2020}.

In the present article, we further explore the consequences of the martingale property 
in the framework of PQ. For Ising spins, the mean equilibrium spin determines completely the probability of the next quenched spin \cite{Note1}. When, furthermore, the mean equilibrium spin is martingale, the Markovian evolution of the total quenched magnetization is found to have a local invariance in its probabilistic path weight.
After the brief description of model and notations in Section \ref{sec:model}, this invariance property is explained in Section \ref{subsec:loc-inv}.

There are two major consequences, both of which were - at least for us -  unexpected and were first recognized through numerical simulations. 
Our first finding is that, given the number of quenched spins $T$, the probability distribution for 
the quenched magnetization $M$ can be expressed as a Boltzmann factor 
containing a ``path-weight potential'' and a ``path-counting entropy'' defined on the $(T,M)$-space.
This result will be described in Section \ref{subsec:pdf-PQ}
and used to describe the bimodality of the distribution of $M$ in Sec. \ref{subsec:bimodal}. Then in Sec.\ref{subsec:canonical} the canonical structures compatible with the long-term memory of the present PQ model are described.
In Section \ref{sec:RQ} we focus on PQ starting from complete thermal equilibrium without constraints. We show that the probability distribution under a given number of quenched spins can also be obtained as the stable limit distribution of the different process that we call Recycled Quenching (RQ).
The latter process consists of the alternative application of single-step unquenching and single-step quenching of randomly chosen quenched spin and unquenched spin, respectively. The detail of RQ is described in Section \ref{subsec:RQ-setup} followed by the analysis of the limit-cycle distribution in Sec. \ref{subsec:RQ-stationary}. 
Finally in Sec. \ref{sec:PQ-RQ}, the connection to PQ through the martingale is given.
Apparently this result challenges our conventional distinction between a diffusing system (described by a parabolic PDE) and a stationary one (described by an elliptic PDE). 

Our results - although based on a particular model - show what the martingale can bring beyond its original definition in terms of the conditional expectation. At the moment it is unknown to what extent our results can be generalized. 
More discussion is given in Section \ref{sec:Discussion}.

\section{Model description, terminology, and short summary of the previous results}\label{sec:model}

\paragraph{Globally coupled spin model:}
We consider the ferromagnetic Ising model on a complete network of $N_0$ spins. Any one of the spins interacts with all the other spins with equal coupling constant, $j/N_0.$ The temperature of the heat bath $T$ is fixed and we absorb $\beta=(\kT)^{-1}$ in $j$. 
It is known that in the limit $N_0\to\infty$ the system undergoes a mean-field phase transition at the critical coupling, $j=j_{crit}(N_0=\infty)=1.$
In order to see clearly the effect of fluctuations, we set the coupling constant $j$ such that the whole system before Progressive Quenching is at the ``critical point '' - i.e the value that maximizes the magnetic susceptibility - of the finite system, $j_{\rm crit}(N_0),$ determined numerically from Curie's law \cite{Note2} (for more details, see \cite{PQ-KS-BV-pre2018}). 

\begin{figure}
	\begin{center}
		\includegraphics[width=0.48\textwidth]{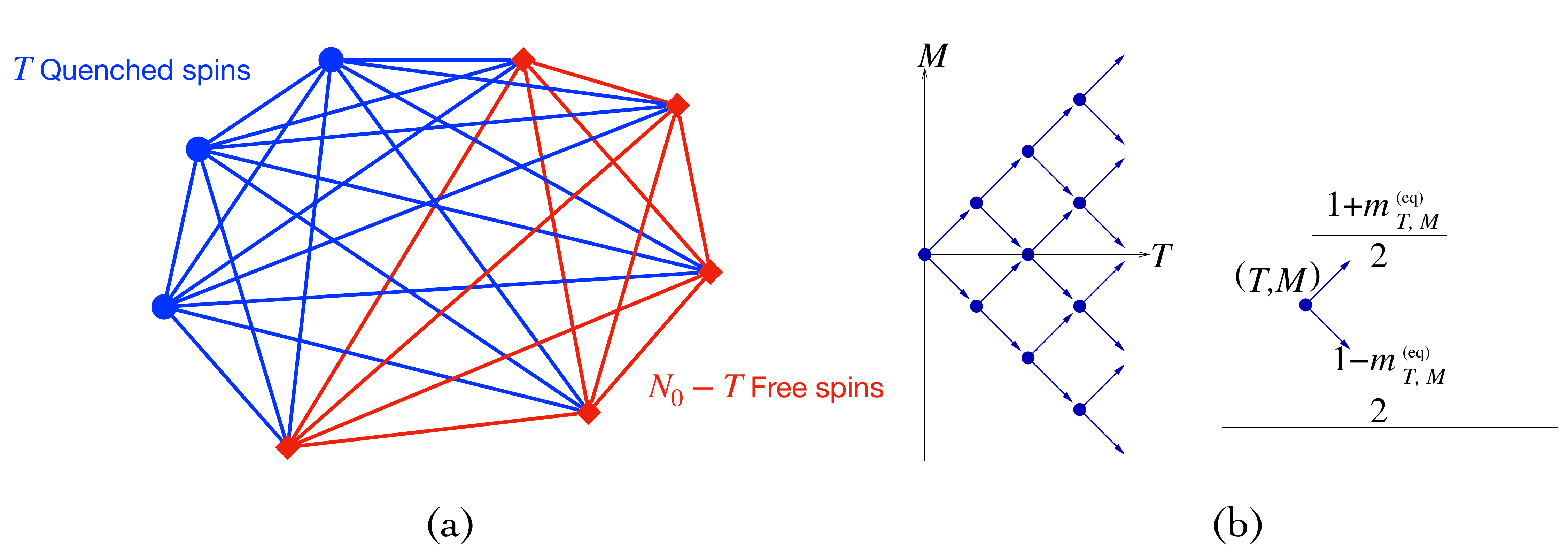}
		\caption{
			(a) In the complete network of $N_0 (=9)$ spins, $T(=3)$ spins have been quenched and there remain $N_0-T(=6)$ free spins. Every spin  interacts with all the other ones by a ferromagnetic coupling constant, $j/N_0.$\\
			(b) PQ is a Markov process representable by a 2D directed network on the integer lattice coordinated by $T$ and $M=M_T=\sum_{k=1}^T s_k.$ If $T$ is even [odd], $M$ is even [odd], respectively.
		}
		\label{fig:intro} 
	\end{center}
\end{figure}

\paragraph{Progressive Quenching (PQ):}
We fix one after another the spins quasi-statically at the value which they took. We call this operation {\it ``quench''}.
We mean by stage-$T$, or simply $T$, that there are $T$  spins that have been quenched, see Fig.\ref{fig:intro}(a) for an illustration.
Fixing quasi-statically means that the interval between the consecutive quenches is large enough for the unquenched spins to reach thermal equilibrium with the heat bath, under the influence of the already quenched $T$ spins which exert a constant external magnetic field  $h=\frac{j}{N_0}M,$ where $M$ is the sum of quenched spins. We shall call $M$ the quenched magnetization for short, and will write $M_T$ when we need to specify the stage $T$.

\paragraph{PQ as Markov process of $M$:}
We denote by $\meq_{T,M}$ the mean equilibrium spin at the stage-$T$ when the quenched magnetization is $M.$ After quenching $(T+1)$-th time, the quenched magnetization $M_T$ is updated by either +1 or –1. See Fig.\ref{fig:intro}b. On the average, it
changes by the average unquenched magnetization, $m^{(eq)}_{T,M_T}$.
Therefore, the expectations values of
the quenched and the unquenched magnetizations are related by
\beq \label{eq:meq-def}
\meq_{T,M_{T}}=E[M_{T+1}-M_{T}| M_{T}]
\eeq
where
$E[ A|B]$ means the conditional expectation of $A$ under the condition $B$.
The quenched magnetization $M_T$ versus $T$ is a Markovian stochastic process if we regard $T$ as the integer {\it time}. 
For the transition from the stage-$T$ to $(T+1),$ probability for the newly quenched spin to be $\pm 1$ is $(1\pm \meq_{T,M})/2,$ respectively. 

\paragraph{Fock-like space of probability distributions:}
The statistical quantity of main interest is the probability distribution of quenched magnetization, $\{P(T,M)\}\equiv \{P(T,-T),P(T,-T+2),\ldots, P(T,T)\}$ at each stage $T$.
Such distribution can be treated as a vector $\vec{P}(T)$ in the $(T+1)$-dimensional Euclidean space. 
Because of the normalization condition
this vector in fact spans a $T$-dimensional simplex.
When we consider the evolution of the probability distribution from $T=0$ 
where $P(0,0)=1,$ up to $T=N_0,$ we effectively use a kind of Fock space in which $\vec{P}(T)$ is found in the $T$-th sector. The process of PQ is a linear mapping between adjacent sectors from, for example $\vec{P}(T)$ to $\vec{P}(T+1)$ through a transfer matrix. 
We have found \cite{PQ-KS-BV-pre2018} that under the critical coupling $j=j_{crit}(N_0)$ the distribution $\vec{P}(T)$ undergoes a unimodal to bimodal transition for some $T$, whose value depends on $N_0.$

\paragraph{``Hidden'' martingale} $\meq_{T,M}$ \textit{:}
Apart from the bimodality of distribution $P(T,M),$ it has been found that $\meq_{T,M}$ is a martingale process induced by the Markovian process $\{ M_T\}$ \cite{PQ-KS-BV-pre2018}. 
The martingale property reads : 
\beq \label{eq:meq-mtg}
E[\meq_{T+1,{M}_{T+1}} |{M}_{T}]=\meq_{T,{M}_{T}}
\eeq 
Eqs.(\ref{eq:meq-mtg}) and (\ref{eq:meq-def}) are the general definition of the ``hidden'' martingale, being independent of the details of the PQ model and the coupling strength $j.$
In our previous studies \cite{PQ-KS-BV-pre2018} we have derived (\ref{eq:meq-mtg}) based on the model represented by 
Fig.\ref{fig:intro}(a) and left a possible finite-size correction of the order of $\mathcal{O}((N_0)^{-2})$.
It turns out that this equality holds exactly, as we show in the Supplemental Material {\bf S1} \cite{SuppMat}.
A more general version of the hidden martingale can thus be formulated: if we define $m_{T,M_{T}}$ by 
$m_{T,M_{T}}\equiv E[M_{T+1}-M_{T}| M_{T},\ldots,M_{0}],$ then its martingale property,
$$E[m_{T+1,M_{T+1}}|M_{T},\ldots,M_0]=m_{T,M_T},$$
holds exactly  with any interactions and evolution dynamics for Ising spins, as long as the spins are homogeneous 
\cite{Note4}. This particular point is detailed in the Section \ref{subsec:canonical} and in the Supplemental Material \textbf{S1} \cite{SuppMat}. Eq.(\ref{eq:meq-mtg}) is thus a special case when $M_T$ is a Markovian process.

\section{Martingale property as a local invariance and its consequence in PQ}
\label{sec:loc-inv}
\subsection{Local invariance of the path weight}\label{subsec:loc-inv}
\begin{figure}
	\centering
	\includegraphics[width=0.46\textwidth]{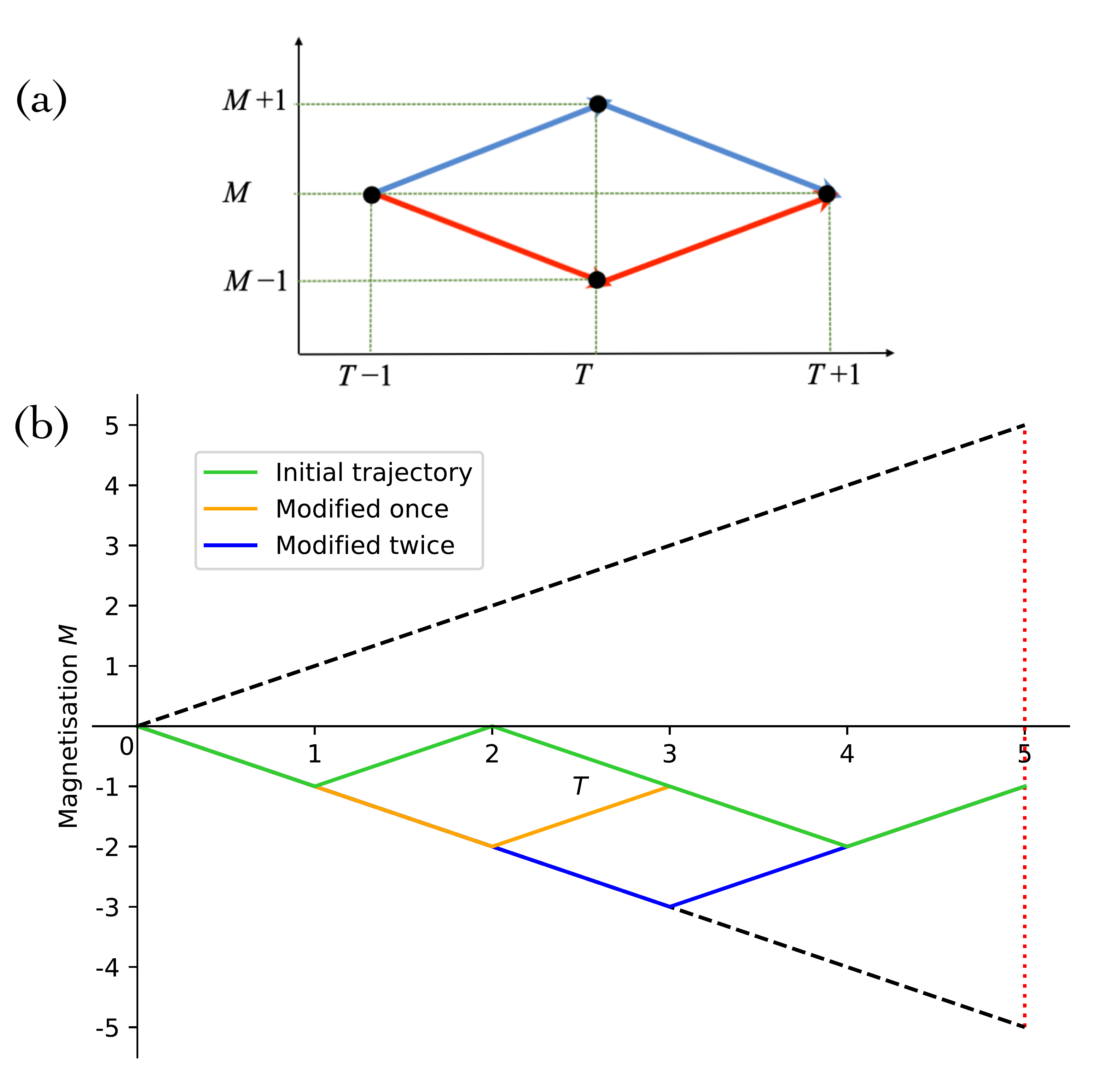}
	\caption{(a): Local invariance of the path weight as a consequence that
		the mean equilibrium spin $\meq_{T,M}$ is martingale.
		The upper (blue) and lower (red) paths are weighted, respectively, by the l.h.s. and r.h.s. of Eq.(\ref{eq:loc-inv}).\\
		(b): Three representative paths connecting $(T,M)=(0,0)$ and $(5,-1).$ All the three paths have the same probability weight due to the local invariance relation Eq.(\ref{eq:loc-inv}).}
	\label{fig:my_label}
\end{figure}

As $M_T-M_{T-1}$ takes the Ising spin variable, the conditional probabilities in (\ref{eq:meq-mtg}) are given in terms of $\meq_{T,M_{T}}$ , and we have
\beq \label{eq:meq-mtg-bis}
\begin{aligned}
	\meq_{T-1,{M}_{T-1}} &=
	\meq_{T,{M}_{T-1}+1} \frac{1+\meq_{T-1,{M}_{T-1}}}{2}\\
	&+\meq_{T,{M}_{T-1}-1} \frac{1-\meq_{T-1,{M}_{T-1}}}{2},
\end{aligned}
\eeq
where, for later convenience, we have shifted the time $T$ by one.
Using the identity $2c-a{(1+c)}{}-b{(1-c)}{}= (1+c)(1-a)-(1-c)(1+b)$, Eq.(\ref{eq:meq-mtg-bis})
can be rewritten  in the form of a {\it local invariance of path-weight} for the stochastic process $M$.
\beq \label{eq:loc-inv}
\begin{aligned}
	&\inRbracket{\frac{1+\meq_{T-1,{M}} }{2}}\inRbracket{\frac{1-\meq_{T,{M}+1} }{2}}\\
	= &\inRbracket{\frac{1-\meq_{T-1,{M}} }{2}}\inRbracket{\frac{1+\meq_{T,{M}-1} }{2}}
\end{aligned}
\eeq
where ${M}_{T-1}$ has been simply denoted by $M.$
Schematically (\ref{eq:loc-inv}) implies that the path weight is invariant under a local change between $(T-1,M)\to(T,M+1)\to(T+1,M)$ and $(T-1,M)\to(T,M-1)\to(T+1,M),$ see Fig.\ref{fig:my_label}(a).

The local invariance shown in Fig.\ref{fig:my_label}(a) reduces significantly the number of independent transition probabilities down to just an extensive one that only depends on the start and end points of the path considered.
In fact, the $\frac{T(T-1)}{2}$ plaquettes \cite{Kogut-RMP} like Fig.\ref{fig:my_label}(a) between $T=0$ ant $T=T$ impose as many constraints on $\meq_{T',M}$ with $0\le T'\le T-1.$ As the latter counts $\frac{T(T+1)}{2}$ values, the difference makes $T$. Moreover, the symmetry with respect to $\pm M$ reduces the freedom 
among $ \{\meq_{M,T} \}$ down to $\lfloor{\frac{T}{2}\rfloor},$ where $\lfloor{ x \rfloor}$ is the floor function. The reduction of independent weight may reflect the persistent memory that we have found before \cite{PQ-CM-KS-2020} .

\subsection{Probability distributions of PQ} \label{subsec:pdf-PQ}

The new property of the martingale $\meq_{T,M}$ in (\ref{eq:loc-inv})
reveals a ``thermodynamic'' structure in the evolution of $\vec{P}(T).$
In general, the probability $P(T,M)$ is the sum of the path weight over all paths arriving at $(T,M)$ from $(0,0).$ 
However, the relation (\ref{eq:loc-inv}) 
in the present system implies the 
degeneracy of all such path weights. For illustration Fig.\ref{fig:my_label}(b) shows
the three paths among those reaching $(T,M)=(5,-1)$ from $(0,0).$
The green (top) path can be represented as a binary sequence, $01001,$ where $1$  $[0]$ means, respectively, to quench $+1$ $[-1]$ spin.
The relation (\ref{eq:loc-inv}) means that the path weight is unchanged 
if we exchange any pair of neighboring bits. Therefore, the orange (middle) path, $00101,$ and then the blue (bottom) path, $00011,$ have the same path weight as the green (top) one.

The immediate consequence is that all the paths connecting the origin
$(0,0)$ to a certain destination $(T,M)$ through PQ have the same weight, which only depends on the number of $1$ [$0$] bits, or equivalently,
on $(T,M),$ see Fig. \ref{fig:my_label}(b). We shall denote such weight by $e^{-\tilde{\beta}\mathcal{E}(T,M)},$ where $\tilde{\beta}\equiv 1$ and the function $\mathcal{E}(T,M)$ gives a ``path-weight potential'' landscape on the $(T,M)$ plane. 
Having known the individual path weight, the sum of the path weight is obtained by counting the number of distinct paths connecting $(0,0)$ and 
$(T,M),$ which is the binomial coefficient $\binom{T}{\frac{M+T}{2}}$. 
We shall denote  this number by $e^\mathcal{S},$ where 
$\mathcal{S}$ represents  a ``path-counting entropy''. The latter is analogous to the conformational entropy of one-dimensional random walk or free polymer chain. If we regard $(T,M)$ as the mesoscopic "state variable" of PQ, the associated microstates (i.e. the paths reaching $(T,M)$) satisfies equipartition.

In summary the probability $P^{ (PQ)}(T,M)$ is given by the  Boltzmann factor of a ``path free energy", $\mathcal{E}-\inv{\tilde{\beta}}\mathcal{S},$ so that 
\eqn{\label{eq:helmholtz}
	P^{(PQ)}(T,M) = e^{\mathcal{S}(T,M)-\tilde{\beta}\mathcal{E}(T,M)},
}   
where
\beq \label{eq:SandE} 
\begin{aligned}
	e^{\mathcal{S}(T,M)} &=
	\binom{T}{\frac{M+T}{2}}\\
	e^{-\tilde{\beta}\mathcal{E}(T,M) }&=
	\prod_{0\le i<(T-M)/2} 
	\inRbracket{\frac{1-m^{(eq)}_{i,i}}{2}}
	\\
	&\times
	\prod_{1\le i \leq (T+M)/2} \inRbracket{\frac{1+m^{(eq)}_{T-i,M-i}}{2}} 
\end{aligned}
\eeq

This is the first of our main results. Remarkably, the structure of Eq.(\ref{eq:SandE})  corresponds to a constrained canonical equilibrium, with identical entropic factors. The latter is calculated in the Section \textbf{S2} of the Supplemental Material \cite{SuppMat}. By this matching, we also have the equality between $\tilde{\beta} \mathcal{E}(T,M)$ and the canonical energy, which justifies our designation. 

\begin{figure}
	\begin{center}
		\includegraphics[width=0.45 \textwidth]{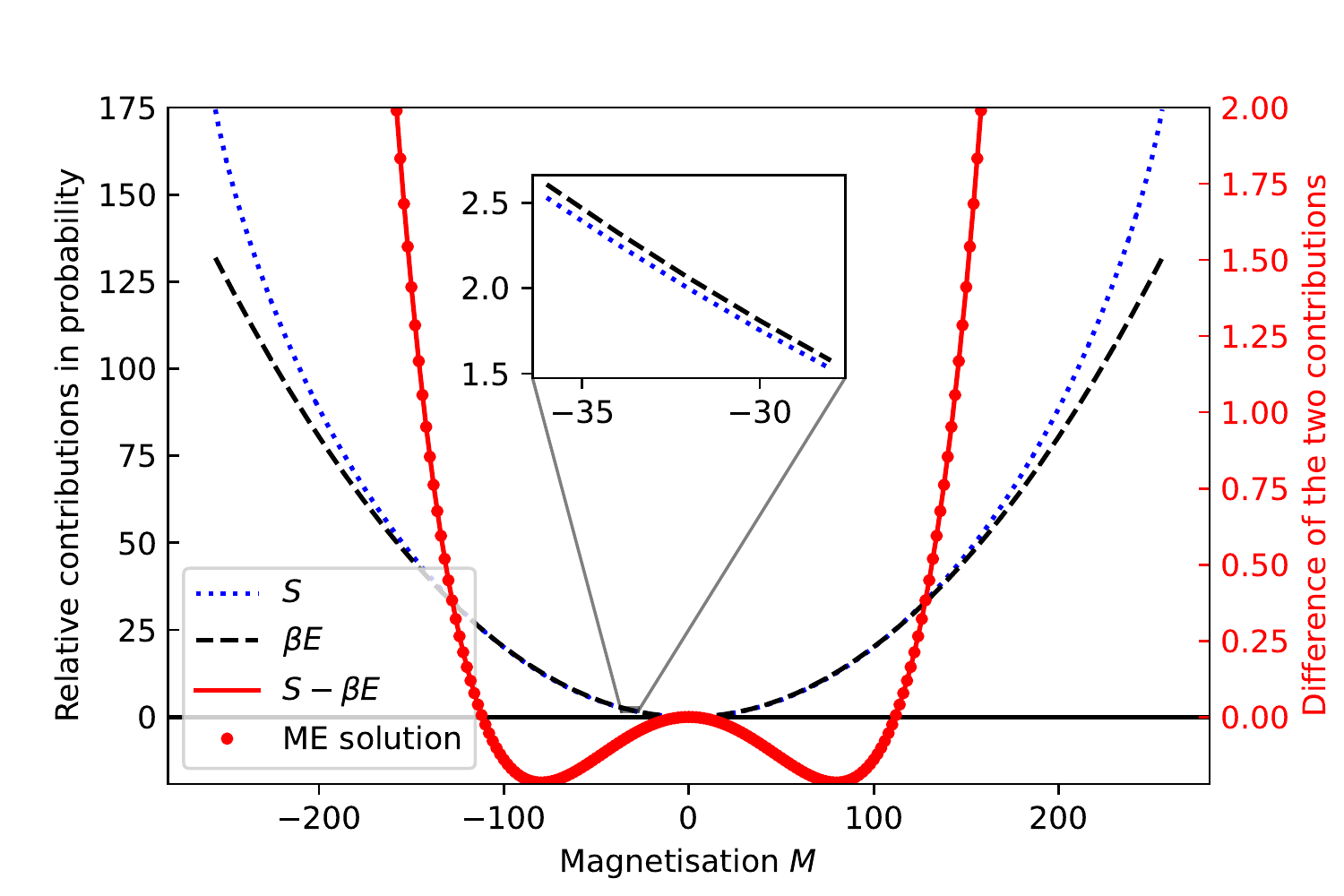}
	\end{center}
	\caption{
		{\it Dashed and dotted curves,  left ordinate}:		 
		$(-\tilde{\beta}\mathcal{E})$ (dashed black) and $(-\mathcal{S})$ (dotted blue) versus $M.$ 
		The value of each curves at $M=0$ is adjusted vertically so that they share the unique origin. These two curves crosses ($\tilde{\beta}\mathcal{E}=\mathcal{S} $) at some finite  $|M|> 0.$
		{\it Solid red curve, right ordinate}:
		The ``path free energy", $\tilde{\beta}\mathcal{E}-\mathcal{S}=(-\log P(T,M))$	($\tilde{\beta}\equiv 1$), versus $M$ (abscissa) for $T=N_0=256$.	
		The large-dotted curve also represents $(-\log P(T,M))$ but directly calculated by the master equation (ME), that is, by the repeated multiplication of the transfer matrix of PQ.}
	\label{fig:contrib}
\end{figure}

In Fig.\ref{fig:contrib} the solid (red) curve shows $\tilde{\beta}\mathcal{E}-\mathcal{S}$ for $T=N_0=256$, while the red-dotted one represents $\log P(T,M)$ which is directly calculated by solving the master equation for the distribution. In Section \ref{sec:PQ-RQ} we will find Eq. (\ref{eq:helmholtz}) by a completely different approach: the ``recycled quenching''.

As a natural extension of the above argument of the path-weight potential and path-counting entropy, we can also have the compact expression of the propagator, $P^{(PQ)}(T, M; T_0, M_0)$ with $0\leq T_0 \leq T \leq N_0,$ which gives the conditional probability for $M_T=M$ to occur at the stage-$T$ given the initial condition
$P^{(PQ)}(T_0,M; T_0,M_0)= \delta_{M, M_0}.$
Following the same argument as (\ref{eq:helmholtz}) and (\ref{eq:SandE})
the value of $P^{(PQ)}(T,M; T_0,M_0)$ can be given in terms of 
$\mathcal{E}(T,M;T_0,M_0)$ and $\mathcal{S}(T,M;T_0,M_0),$ whose detailed account may not be necessary to repeat. 

\subsection{Origin of the bimodality as ``potential-entropy'' trade-off}\label{subsec:bimodal} 
We have encountered bimodal distributions for $M$ during PQ 
even if the coupling $j/N_0$ is not in the ferromagnetic regime. The symmetry breaking does not occur for a finite size $N_0,$ and the propensity of non-zero $M$ should not be taken as the equilibrium phase transition. The above ``thermodynamic''  decomposition allows us to understand how the bimodality of the probability distribution can arise.
We may constitute the following qualitative argument: When the total magnetization $M$ is non-zero, 
the molecular field, $(j/N_0)M,$ on the unquenched spins makes non-zero mean equilibrium spin, $m^{(eq)}_{T,M}.$ This causes the biased probability of subsequently quenched spin, which in turn reinforces the non-zero magnetization $M$ as positive feedback.
This is the scenario for the instability of $\tilde{\beta}\mathcal{E}-\mathcal{S}$ around $M=0.$ 
By contrast, the path-counting entropy factor becomes highly diminished for  $|M| \sim T,$ reflecting the limited availability of paths. This explains the high rise of $\tilde{\beta}\mathcal{E}-\mathcal{S}$ for $|M|\sim T.$
The competition of these two factors can give rise to the bimodal distribution.
At the early stages, $T\ll N_0,$ however, the entropy factor prevails and the distribution is unimodal \cite{PQ-CM-KS-2020} .

While the above ``thermodynamic'' picture explains a qualitative origin of 
bimodality, more subtle question would be whether such aspect persists in the limit of large system, $N_0\to\infty,$ especially when $j$ is chosen to be at the extrapolated Curie point \cite{PQ-KS-BV-pre2018}.
Leaving the detailed account in Section \ref{sec:PQ-RQ} and Supplemental Material {\bf S3} \cite{SuppMat},
the short answer is affirmative and we expect that $P(T=N_0;M)$ has maxima at $M=\pm M^\circ (N_0),$ where $M^\circ (N_0)\sim (N_0)^{1-\frac{\nu}{2}}$ with 
$\nu\simeq 0.933$ being the finite-size scaling exponent such that $j_{crit}(N_0)= 1+c (N_0)^{-\nu}$ \cite{PQ-KS-BV-pre2018}.

\subsection{Constrained canonical statistics by PQ}\label{subsec:canonical}

One might wonder if the equilibrium canonical distribution lies behind the ``thermodynamic'' structure of
(\ref{eq:helmholtz}). The answer is yes but under constraints:
If, and only if, the PQ starts from the unbiased initial condition, 
$M_{T=0}=0$ with probability one, does the probability $P^{(PQ)}(T,M)$ have the canonical equilibrium weight for the event that the group of spins $\{s_1,\ldots,s_T\}$ has the magnetization $\sum_{i=1}^T s_i=M.$ In the Supplemental Material {\bf S2} \cite{SuppMat} we detailed the expression of this canonical equilibrium weight.
By contrast, if $T_0$ spins have already been quenched with their magnetization being $M_0,$ the later probability for $T\ge T_0$, or the propagator $P^{(PQ)}(T,M; T_0,M_0)$ of Section \ref{subsec:pdf-PQ} retains a persistent memory and the distribution coincides with a {\it constrained} canonical weight for the event that the group of spins $\{s_1,\ldots,s_T\}$ has the magnetization, $\sum_{i=1}^T s_i=M,$ {\it under the constraint} that  its subset, $\{s_{i_1},\ldots,s_{i_{T_0}}\} \subset \{s_1,\ldots,s_T\},$ has the magnetization $M_0.$ The fact that this function has a strict support (of causality) $|M-M_0|> T-T_0$ along the $M$ axis is consistent with the above constraint.

Altogether, the two facets of PQ, the neutrality of quenching hitherto equilibrated spins on the one hand, and the persistence of memory in quenched spins on the other hand, are made compatible in the form of the constrained canonical distribution.

Below we argue that the mechanism behind this compatibility is the close relationship between the conditional probability and the act of quenching a spin.
Let us denote by $\Peq(s_i |s_{i-1},\ldots s_1)$ the conditional probability that the $i$-th spin $s_i$ takes the specified value ($\pm 1$) in a canonical equilibrium ensemble of $N_0$ spins, $\{s_1,\ldots,s_{N_0}\},$ given that the spins $\{s_1,\ldots, s_{i-1}\}$ are found to take the specified values. 
Also let us denote by $\Ppq(s_i |s_{i-1},\ldots s_1)$  the conditional probability that the $i$-th spin $s_i$ takes the specified value ($\pm 1$) upon quenching in a {\it constrained} canonical equilibrium ensemble of $N_0-(i-1)$ spins, $\{s_i,\ldots,s_{N_0}\},$ given that the other spins $\{s_1,\ldots, s_{i-1}\}$ have already been frozen to take the specified values. 
We may then expect the following equality, 
\beq \label{eq:pqeq2}
\Ppq(s_i |s_{i-1},\ldots s_1)= \Peq(s_i |s_{i-1},\ldots s_1).
\eeq
On the other hand, if the first spin $s_1$ has been quenched when the whole system $\{s_1,\ldots,s_{N_0}\}$ was in equilibrium, the probability of the  quenched spin $\Ppq(s_1)$ should be equal to the equilibrium one:
\beq \label{eq:pqeq1}
\Ppq(s_1)=\Peq(s_1).
\eeq
We then have the equality of the joint probabilities, 
\beq \label{eq:pqeq3}
\Ppq(s_n, \ldots s_1)= \Peq(s_n,\ldots s_1),
\eeq
for $2\le n\le N_0$ because of the general chain rule which is valid for both $\Ppq$ and $\Peq$ :
\beqa
P(s_n, \ldots s_1)&=& \prod_{i=2}^{n} \inRbracket{\frac{P(s_i,\ldots,s_1)}{P(s_{i-1},\ldots,s_1)}} P(s_1)
\cr &=&
\inRbracket{\prod_{i=2}^{n} {P(s_i| s_{i-1},\ldots,s_1)} } P(s_1).
\nonumber
\eeqa
Said differently, freezing spins one by one quasi-statically gives the same result as freezing all of them altogether as a snapshot.

While (\ref{eq:pqeq2}) seems to hold for the quasi-equilibrium quenching
with any choice of $\{s_{i-1},\ldots,s_1\},$ the last result (\ref{eq:pqeq3}) holds
only with the equilibrium starting point (\ref{eq:pqeq1}). 
If the PQ starts from $\Ppq(s_1)$ other than $\Peq(s_1)$ or 
from some prefixed spins $\{s_{n_0},\ldots, s_1\},$ the progression of PQ carries non-volatile memory preventing the relaxation to the canonical weight.

\section{Recycled Quenching (RQ)}\label{sec:RQ}
\subsection{Single-step unquenching \texorpdfstring{$\bm{S}$}{S} and 
	single-step quenching \texorpdfstring{$\bm{K}$}{K} } 
\label{subsec:RQ-setup}

\begin{figure*}
	\begin{center}
		\includegraphics[width=1.0\textwidth]{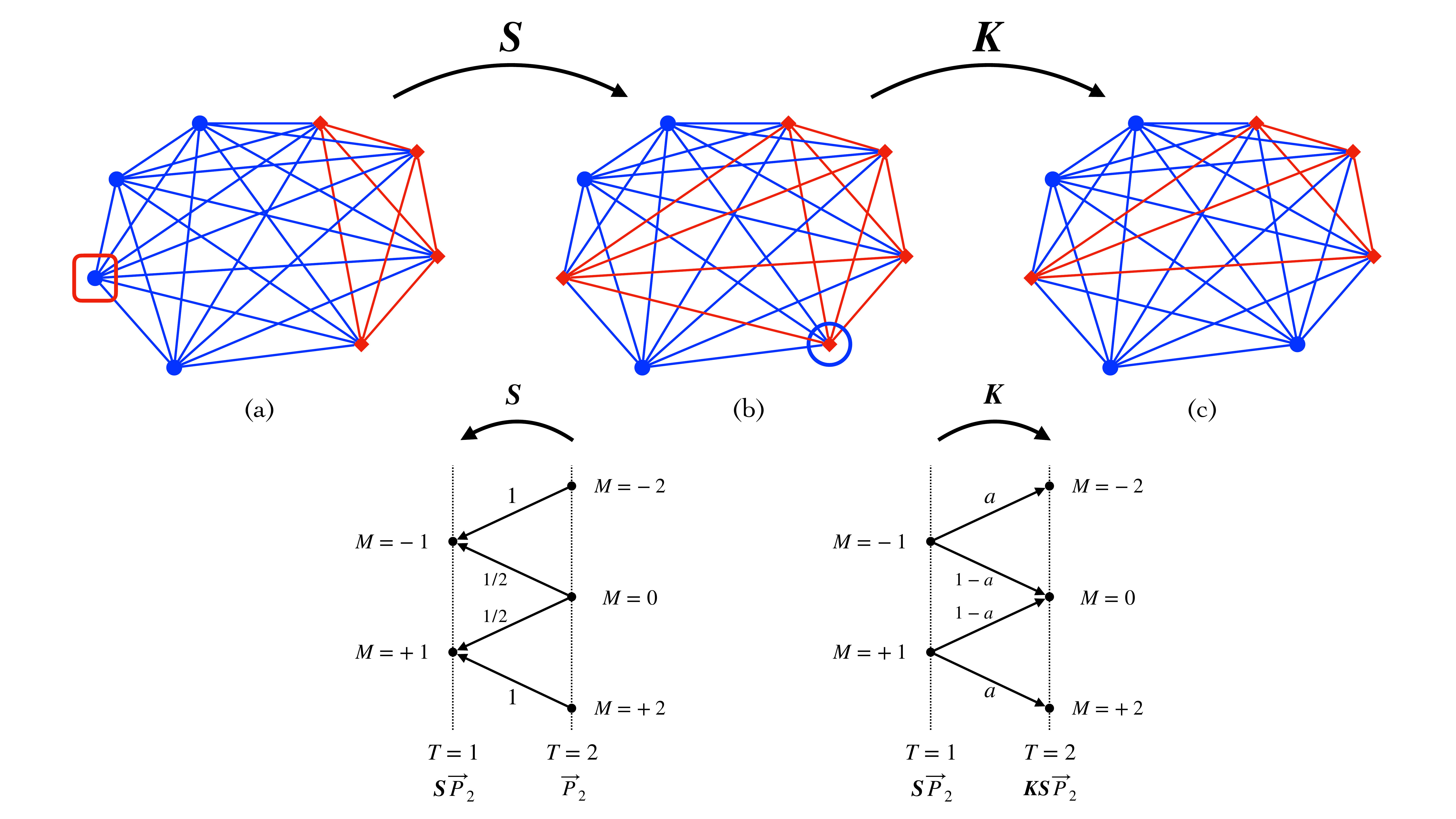}
		\caption{
			{\it Top row:} Schematic representation of the recycled quenching process.
			(a): Step $\bm{S}$ : A quenched spin - blue circle (darker gray) and squared in red (lighter gray) - is picked at random and is un-quenched. 
			(b): Step $\bm{K}$: A unquenched spin (red square circled in blue) is quenched as in the Progressive Quenching.
			(c): Updated state of the system after operating $\bm{S}.$ then $\bm{K}.$
			\textit{Bottom row:}  Probability tree of the operation of $\bm{S}$ {\it (left)} 
			and $\bm{K}$ {\it (right)} 
			over a distributions for the stages $T=1$ and $T=2$.
		}
		\label{fig:RQprotocol}
	\end{center}
	
\end{figure*}

Let us leave momentarily from the analysis of progressive operation of quenching (PQ)
and rather consider the cyclic operation of a single-step quenching and un-quenching (recycled quench, or RQ for short). 
See Fig.\ref{fig:RQprotocol}.
We propose the following process : Take again a system of $N_0$ Ising spins on a complete network as being specified in Section \ref{sec:model}.
Suppose $T$ spins are quenched with a total quenched magnetization $M$ while the $N_0-T$ remaining spins are thermalized with a bath. We then select at random a quenched spin and allow it to be un-quenched (operation $\bm{S}$). Subsequently, after reaching thermal equilibrium once again , we apply a single step of quenching step as in Sec. \ref{sec:model} and \cite{PQ-KS-BV-pre2018}\cite{PQ-CM-KS-2020} (operation $\bm{K}$).
While the number of quenched spin returns from $T-1$ to $T,$ the updated state of the system may have its quenched magnetization either set to $M$ or $M\pm 2$.

By applying alternatively the unquenching ($\bm{S}$) and 
quenching ($\bm{K}$) we generate a series of probability distributions, which may be written as follows: 
\beq \label{eq:S-K-sequence}
\stackrel{\bm{S}}{\rightarrow} \vec{Q}^{[\ell]}(T-1)
\stackrel{\bm{K}}{\rightarrow} \vec{P}^{[\ell]}(T)
\stackrel{\bm{S}}{\rightarrow} \vec{Q}^{[\ell+1]}(T-1)
\stackrel{\bm{K}}{\rightarrow} \vec{P}^{[\ell+1]}(T)
\stackrel{\bm{S}}{\rightarrow}
\eeq
where $\vec{P}$ and $\vec{Q}$ denote the probability vectors of having a certain magnetization after a step $K$ or $S$, respectively, and the superfix $[\ell]$ etc. merely counts the number of iterated operations, and the number of fixed spins, $T,$ is no more the `time'.

\begin{figure}
	\begin{center}		
		\includegraphics[width=0.35\textwidth]{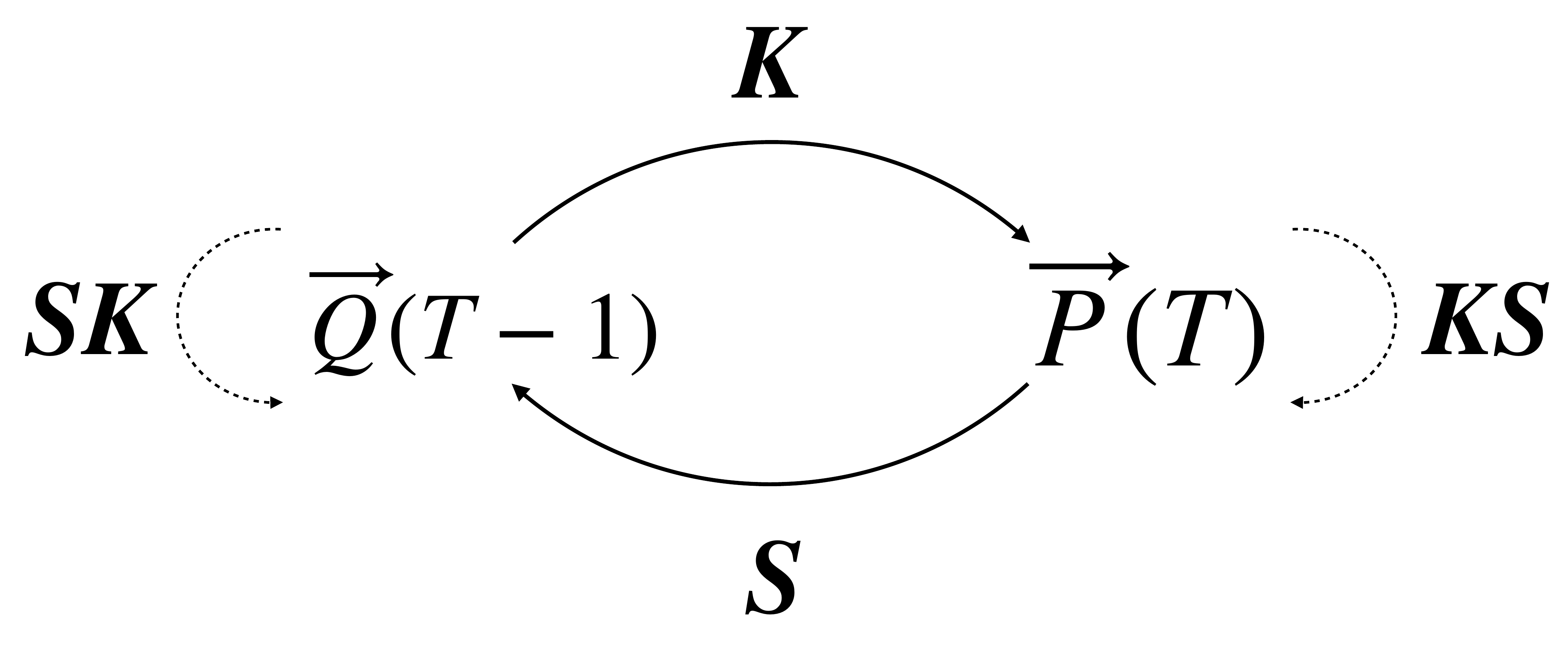}
	\end{center}
	\caption{Symbolic representation of the action of the recycling operators over the distributions $\vec{P}(T)$ and $\vec{Q}(T-1)$.}
	\label{fig:symbl}
\end{figure}

If we focus on $\vec{P}(T)$'s, a single application of this \textit{recycling} process can be seen as transformation over the probability vector $\vec{P}(T)$ by two operators : $\bm{S}$ then $\bm{K},$  
leading to 
\beq \label{eq:KSP} 
\vec{P}^{[\ell+1]}(T)=\bm{(KS)}\vec{P}^{[\ell]}(T).
\eeq
Alternatively, if we focus on $\vec{Q}(T-1)$'s, we can think of a adjoint process, where the two steps are reversed in order, i.e. $\bm{K}$ then $\bm{S},$ 
leading to 
\beq \label{eq:SKQ} 
\vec{Q}^{[\ell+1]}(T-1)=\bm{(SK)}\vec{Q}^{[\ell]}(T-1).
\eeq
In either point of view the recycling process retains the number of quenched spins. Altogether we can schematize the operation of unquenching and quenching in the form of Fig.\ref{fig:symbl}.
The detailed action of  $\bm{K}$ and $\bm{S}$ over a probability distribution is accounted in the Supplemental Material, Section {\bf S4} \cite{SuppMat}.

\subsection{Stationary distributions}\label{subsec:RQ-stationary}
\paragraph{Case studies:} 
Because the number of quenched spins remains the same after the action of $\bm{KS}$ and $\bm{SK}$, these combined operations
are the transfer matrix on the vectors $\vec{P}$ and $\vec{Q},$ respectively. Applying the Perron-Frobenius theorem to those matrices ensures the existence of the non-degenerate maximum eigenvalue which is unity. Thus, we expect the presence of unique {\it stable} stationary distributions,
$\vec{P}^{[\infty]}(T)$ and $\vec{Q}^{[\infty]}(T-1),$ respectively.
To understand intuitively the stability or convergence, we consider the cases $T=2$ and $3$ below.

$\bm{T=2}$ \textbf{case:}
Fig.\ref{fig:RQprotocol} (bottom left)
indicates the transfer probabilities assigned to $\bm{S}$ acting on $\vec{P}_2,$ where $\vec{P}_2=(P(2,-2),P(2,0),P(2,+2))^t,$ and 
Fig.\ref{fig:RQprotocol} (bottom right) indicates the transfer probabilities assigned to $\bm{K}$ acting on $\bm{S}\vec{P}_2,$ where 
$a \equiv  \frac{1+m^{(eq)}_{1,+1}}{2} =\frac{1-m^{(eq)}_{1,-1}}{2}$.
The transfer matrix $\bm{KS}$ is, in this case :
\[\bm{KS} = \begin{pmatrix} 
	a & \frac{a}{2} & 0 \\
	1-a & 1-a & 1-a \\
	0 & \frac{a}{2} & a
\end{pmatrix}  \]
A simple induction gives an explicit formula for $\bm{(KS)}^N $ and 
its convergence: 
\beq
\begin{aligned}
	\bm{(KS)}^N 
	&= \begin{pmatrix} 
		\frac{a}{2} + \frac{a^N}{2}  & \frac{a}{2} & \frac{a}{2} - \frac{a^N}{2} \\
		1-a & 1-a & 1-a \\
		\frac{a}{2} - \frac{a^N}{2} & \frac{a}{2} & \frac{a}{2} + \frac{a^N}{2}
	\end{pmatrix}\\
	&\stackrel{N \rightarrow \infty }{\longrightarrow}
	\begin{pmatrix} 
		\frac{a}{2} & \frac{a}{2} & \frac{a}{2} \\
		1-a & 1-a & 1-a \\
		\frac{a}{2} & \frac{a}{2} & \frac{a}{2}
	\end{pmatrix}  
\end{aligned}
\eeq
Therefore, from whatsoever distribution $\vec{P}_2$ the result of RQ cycle,
$\bm{(KS)}^N \vec{P}(2),$ converges to the stationary distribution: 
$\vec{P}^{[\infty]} (2)
=(\frac{a}{2},1-a,	\frac{a}{2})^t.$ 
We notice that this stationary distribution coincides with the one obtained by the progressive quenching from ${P}^{ (PQ)}(0,0)=1,$ that is
$\vec{P}^{ (PQ)}(2) =\vec{P}^{[\infty]}(2)$ (see below). 

$\bm{T=3}$ \textbf{case :}
We can make  the scheme  similar to Fig.\ref{fig:RQprotocol} (bottom) to find the transfer matrix $\bm{KS}$. We then obtain :
$$ \bm{KS} = \begin{pmatrix} 
	b & \frac{b}{3} & 0 & 0\\
	1-b & \frac{2-b}{3} & \frac{1}{3} & 0 \\
	0 & \frac{1}{3} & \frac{2-b}{3} & 1-b \\
	0 & 0 & \frac{b}{3} & b
\end{pmatrix}, $$ 
where $b \equiv \frac{1+m^{(eq)}_{2,+2}}{2}=\frac{1-m^{(eq)}_{2,-2}}{2}.$
Expression for $\bm{(KS)}^N$ 
is rather cumbersome but we know the convergence of $\bm{(KS)}$ by its eigenspectrum, $\{1,\frac{2b+1}{3},\frac{2b}{3}, 0\},$ where we have
$1>(2b+1)/3>2b/3>0$ because $0<b<1.$
The normalized eigenvector corresponding to the steady state is :  $\vec{P}^{[\infty]}(3) = 
(\frac{b}{2(3-2b)},\frac{3(1-b)}{2(3-2b)},\frac{3(1-b)}{2(3-2b)},\frac{b}{2(3-2b)})^t .$ 
To compare, the distribution obtained by the progressive quenching
reads $\vec{P}^{PQ}(3) = (	\frac{ab}{2},	\frac{1-ab}{2},	\frac{1-ab}{2} ,	\frac{ab}{2})^t$  with $b$ just defined and $a \equiv \frac{1+m^{(eq)}_{1,1}}{2}$ already defined above. This apparently different distribution is in fact identical to the former, $\vec{P}^{[\infty]}(3),$ because the martingale (\ref{eq:meq-mtg-bis}) 
- or the local invariance (\ref{eq:loc-inv}) - imposes the relation, $a=\inv{3-2b}.$\\

\paragraph{General case :}

Altogether, from the previous case studies 
we admit that the iterative operation of $\bm{(KS)}$
or $\bm{(SK)}$ on a probability vector of the $T$-sector
brings about the convergence to $\vec{P}^{[\infty]}(T)$ and $\vec{Q}^{[\infty]}(T),$ respectively, as stable fixed points:
\beqa \label{eq:FP}
\bm{(KS)}\vec{P}^{[\infty]}(T)&=& \vec{P}^{[\infty]}(T)
\cr 
\bm{(SK)}\vec{Q}^{[\infty]}(T)&=&\vec{Q}^{[\infty]}(T)
\eeqa
These fixed points are also the eigenvectors of these operators with the maximum eigenvalue ($=1$).
Using the concrete expressions for the action of $\bm{(KS)}$ and $\bm{(SK)}$
in Supplemental Material {\bf S4} \cite{SuppMat},
the equations in (\ref{eq:FP}) can be rewritten as follows, where we use the notations, $p_M=P^{[\infty]}(T,M)$ and  $q_M=Q^{[\infty]}(T,M)$:
\begin{equation}
	\label{eq:KS-fp}
	\begin{aligned} 
		0&=p_{M-2}\inRbracket{1-\frac{M-2}{T}}\inRbracket{1+ \meq_{T-1,M-1}} \\
		&-p_{M}\inRbracket{1+\frac{M}{T}}\inRbracket{1- \meq_{T-1,M-1}} \\
		&-\left[ p_{M}\inRbracket{1-\frac{M}{T}} \inRbracket{1+ \meq_{T-1,M+1}} \right. \\
		&\left.-p_{M+2}\inRbracket{1+\frac{M+2}{T}}\inRbracket{1 -\meq_{T-1,M+1}}\right]
	\end{aligned}
\end{equation}
and similarly : 
\begin{equation}
	\label{eq:SK-fp}
	\begin{aligned} 
		0&=q_{M-2}\inRbracket{1-\frac{M-1}{T+1}}\inRbracket{1+ \meq_{T,M-2}} \\
		&-q_{M}\inRbracket{1+\frac{M-1}{T+1}}\inRbracket{1- \meq_{T,M}} \\
		&-\left[q_{M}\inRbracket{1-\frac{M+1}{T+1}}\inRbracket{1+ \meq_{T,M}}\right. \\
		&\left.-q_{M+2}\inRbracket{1+\frac{M+1}{T+1}}\inRbracket{1 -\meq_{T,M+2}} \right]
	\end{aligned}
\end{equation}
Since $[\cdots]$ in the second lines are simply shifted by $+2$ for the variable $M$ with respect to the first lines,  the ``first integrals" are
\begin{equation*}
	\begin{aligned}
		&p_{M}\inRbracket{1-\frac{M}{T}}\inRbracket{1+ \meq_{T-1,M+1}} \\
		-&p_{M+2}\inRbracket{1+\frac{M+2}{T}}\inRbracket{1 -\meq_{T-1,M+1}}=c_+
	\end{aligned}
\end{equation*}
and
\begin{equation*}
	\begin{aligned}
		&q_{M}\inRbracket{1-\frac{M+1}{T+1}}\inRbracket{1+ \meq_{T,M}} \\
		-&q_{M+2}\inRbracket{1+\frac{M+1}{T+1}}\inRbracket{1 -\meq_{T,M+2}}=c_-,
	\end{aligned}
\end{equation*}
where $c_\pm$ are independent of $T.$
Moreover, it is only for $c_+=0$ [$c_-=0$] that  $p_{T+2}$ [$q_{T+2}$] or $p_{-T-2}$ [$q_{-T-2}$] are not generated. Therefore, $c_\pm=0.$
We then have 
\beq \label{eq:SS-KS}
\frac{p_{M+2}}{p_{M}}= \frac{ \inRbracket{1-\frac{M}{T}}\inRbracket{1+ \meq_{T-1,M+1}} }{\inRbracket{1+\frac{M+2}{T}}\inRbracket{1 -\meq_{T-1,M+1}}} 
\eeq
and 
\beq \label{eq:SS-SK}
\frac{q_{M+2}}{q_{M}}= \frac{ \inRbracket{1-\frac{M+1}{T+1}}\inRbracket{1+ \meq_{T,M}}
}{ \inRbracket{1+\frac{M+1}{T+1}}\inRbracket{1 -\meq_{T,M+2}} }.
\eeq
With the aid of the normalization conditions, the iterative conditions 
(\ref{eq:SS-KS}) and (\ref{eq:SS-SK}) 
should give the stationary distributions 
$\vec{P}^{[\infty]}(T)$ and $\vec{Q}^{[\infty]}(T),$ 
respectively.
\subsection{Martingale connects stationary distributions of RQ to PQ}
\label{sec:PQ-RQ}

\begin{figure}
	\begin{center}
		\includegraphics[width=0.45\textwidth]{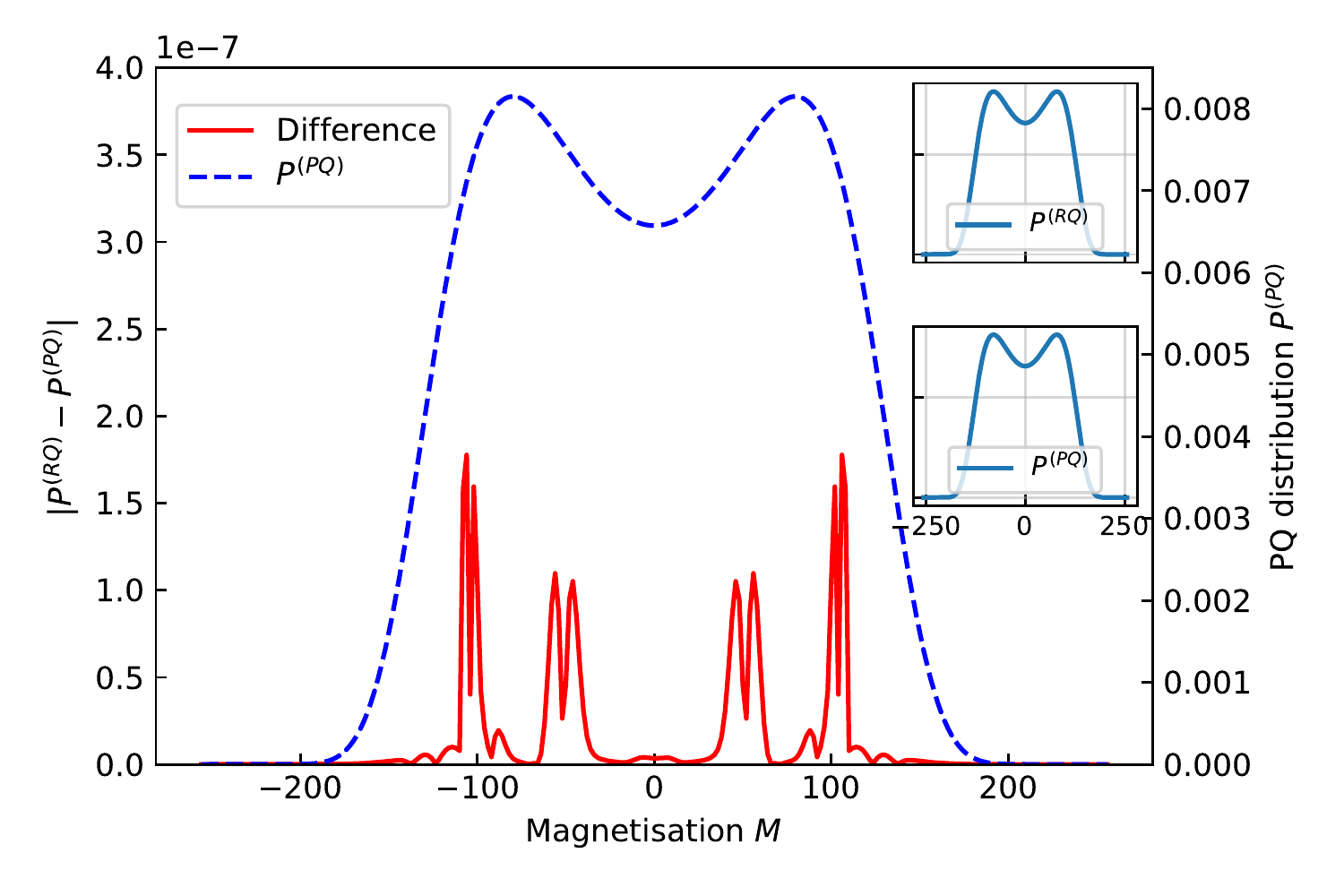}
	\end{center}
	\caption{For $N_0=T=256$ the distribution of PQ, $\vec{P}^{ (PQ)}_T$ {\it (lower inset)} and the stationary distribution of RQ, $\vec{P}^{[\infty]}(T)$ {\it (upper inset)} are compared ({\it solid curve and the left ordinate} in unit of $10^{-7}$). The {\it dashed curve and the right ordinate} shows $\vec{P}^{ (PQ)}_T.$ 
		\label{fig:comparison}}
\end{figure}

\paragraph{Numerical comparisons}:
Having characterized  $\vec{P}^{[\infty]}(T)$ and $\vec{Q}^{[\infty]}(T)$ with any value of $T$ as the stable fixed distributions of $\bm{(KS)}$ and 
$\bm{(SK)},$ respectively, we evaluated numerically these distributions
for different $T$ and for $N_0.$ It is done by seeking the eigenvectors
corresponding to the largest eigenvalue $(=1).$
To our surprise, our analysis shows that the two stationary distributions, $\vec{P}^{[\infty]}(T)$ and $\vec{Q}^{[\infty]}(T),$  
are extremely similar, and that the similitude increases with the number of spins in the entire system $N_0$. 
Moreover, they are also almost identical to the distribution of the Progressive Quenching, $\vec{P}^{ (PQ)}_T, $ when $N_0\gg 1.$
Fig.\ref{fig:comparison} shows the comparison 
between $\vec{P}^{ (RQ)}(T)\equiv \vec{P}^{[\infty]}(T)$ {\it (upper inset)} and $\vec{P}^{ (PQ)}(T)$ {\it (lower inset)}. The difference of order $10^{-7}$ {\it (solid curve in red)} is much smaller than the probability distribution, which is of order $10^{-2}$ {\it (dashed curve in blue)} in the case of $N_0=T=256.$

\begin{figure}
	\begin{center}
		\includegraphics[width=0.45 \textwidth]{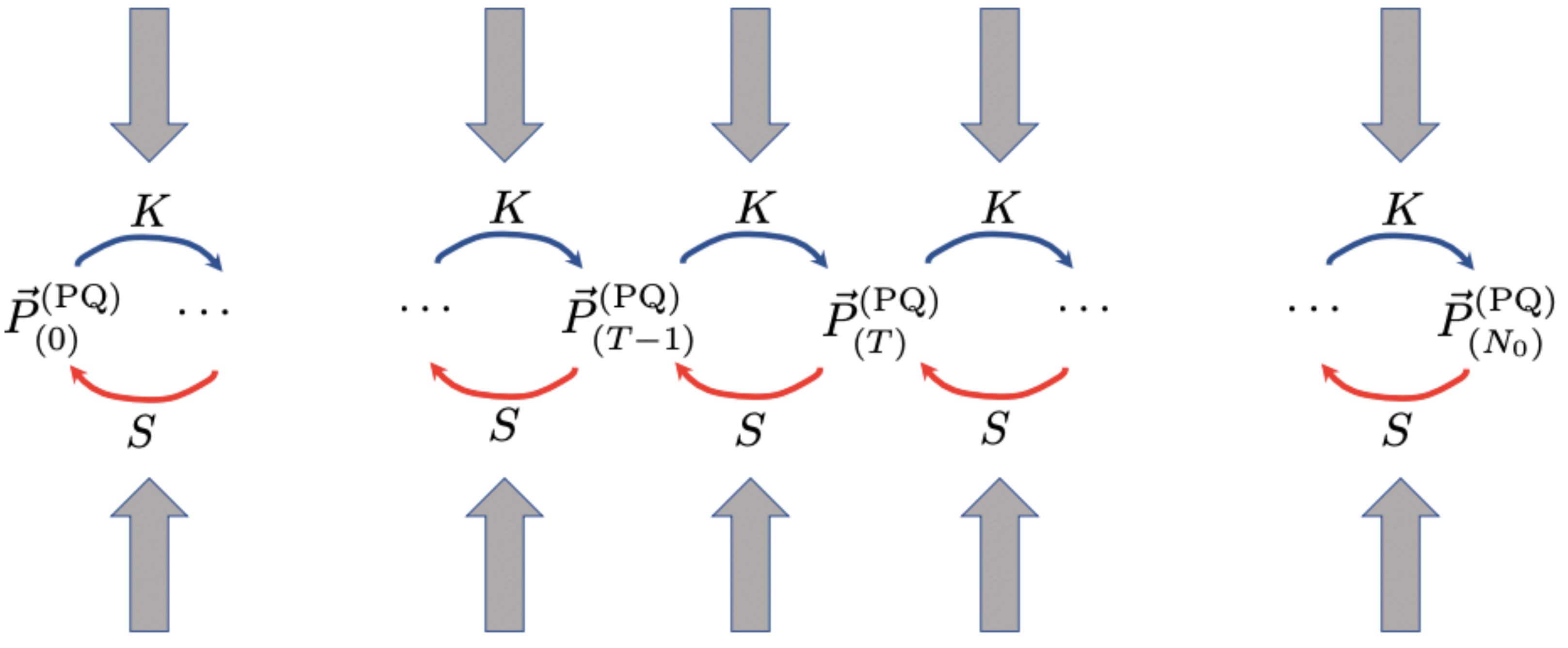}
	\end{center}
	\caption{While the progressive quenching (the symbol $\bm{K}$ and blue arrows) generates $\vec{P}^{ (PQ)}(T)$ from $\vec{P}^{ (PQ)}(T-1),$ 
		the random unquenching of quenched spins (the symbol $\bm{S}$ and red arrows) generates $\vec{P}^{ (PQ)}(T-1)$ 
		from $\vec{P}^{ (PQ)}(T)$
		as the ``on-shell'' reverse operation.
		At the same time, the family of these distributions $\{\vec{P}^{ (PQ)}(T)\}_{T=0}^{N_0}$ are the attractor of the Recycling Quenching, $\bm{KS}$ {\it and} $\bm{SK}$ (the upward and downward thick arrows).
		\label{fig:seriesKS}}
\end{figure}

\paragraph{Implication of martingale}: 
The key to understand the above mentioned ``coincidence'' is the martingale.
In fact the local invariance (\ref{eq:loc-inv}), which is equivalent to the martingale property of $\meq_{T,\hat{M}_T},$ Eq.(\ref{eq:meq-mtg}), assures that the r.h.s. of (\ref{eq:SS-KS}) and that of (\ref{eq:SS-SK}) are the same.
To show this we have also used the identity, $\inRbracket{1-  \frac{M}{T}}/\inRbracket{1+  \frac{M+2}{T}}= \inRbracket{1-  \frac{M+1}{T+1}}/\inRbracket{1+  \frac{M+1}{T+1}}.$ 
Under the normalization condition, these two equations, therefore, defines the unique distribution: $\vec{P}^{[\infty]}(T)= \vec{Q}^{[\infty]}(T).$
The consequence of this equality is profound if we recall (\ref{eq:S-K-sequence}) with $\ell=\infty,$ because the latter implies
\begin{align}
	&\bm{K}\vec{P}^{[\infty]}(T-1) = \vec{P}^{[\infty]}(T) \label{eq:twistK} \\
	&\bm{S}\vec{P}^{[\infty]}(T) = \vec{P}^{[\infty]}(T-1) \label{eq:twistS}
\end{align}
Eq.(\ref{eq:twistK}) tells in fact that the whole family of stationary distributions of Recycled Quenching, $\{ \vec{P}^{[\infty]}\}_{T=0}^{N_0},$
is generated by the Progressive Quenching one after another starting from 
the initial one, $\vec{P}^{(PQ)}(0)=1.$ 
\begin{equation}
\vec{P}^{[\infty]}(T)= \vec{Q}^{[\infty]}(T) = \vec{P}^{(PQ)}(T). 
\end{equation}
This is the second of our main results.
This fact, a kind of {envelope relation}, can be also verified by
directly ``integrating'' (\ref{eq:SS-KS}) and comparing with (\ref{eq:helmholtz})
and (\ref{eq:SandE}) (the details not shown).
Eq.(\ref{eq:twistS}) tells that 
the random unquenching of a spin by $\bm{S}$ allows to step back 
the distribution of the Progressive Quenching. 
Schematically we may represent these by Fig.\ref{fig:seriesKS}.

We note that this is ``on-shell'' property, which concerns only the stationary distributions of RQ. In the sense ``off-shell,'' the family $\{\vec{P}^{ (PQ)}_T\}_{T=0}^{N_0}$ constitutes a set of stable attractors  of the RQ operations,
$\bm{KS}$ and $\bm{SK}.$ 

\section{Conclusion and Discussion}\label{sec:Discussion}

In this paper, we showed that the PQ
process has a local invariance induced by the hidden martingale.
This new symmetry allowed us to derive an exact probability formula, which corresponds to the canonical one under unbiased conditions. By introducing a new operation: the single-spin unquenching, we described a new stochastic process - the Recycled Quenching - whose stable stationary distribution is associated to the PQ through the local invariance.

Progressive Quenching, though the operator $\bm{K}$, is an operation by which the partition between the system and its environment is updated, while the unquenching, through $\bm{S}$, is a kind of its inverse. 
In our model, this dichotomy between the system (here, the unquenched spins) and the environment (the quenched ones) subsystems is explicitly made. The quenching operation drives a spin in an out-of-equilibrium state, while the unquenched part remains at equilibrium under the updated constraint.
Such a flexibility of partition opens a niche where we may find new concepts.
The evolution of Progressive Quenching from an unbiased initial condition generates the family of stable steady states for the Recycled Quenching process, the alternation of single-step quenching ($\bm{K}$) and single-step unquenching of a randomly chosen spin ($\bm{S}$). That family of steady distributions plays the role of a stable manifold in the space of distributions with multi-sectors. 

There are several questions that we have not exploited and left for the future study.  We have not addressed the kinetic aspects of RQ, which might bring more information about this new realm of flexible System-Environment partition. As for the PQ, while the canonicality was separately explained in Sec.\ref{subsec:canonical}, we don't fully understand how the hidden martingale (\ref{eq:meq-mtg-bis}) could bring the canonical or Boltzmann-like structure (\ref{eq:helmholtz}) without reference to the canonicality of the unquenched spins but only using the Markovian and Ising characters of quenched spins leading to the local invariance (\ref{eq:loc-inv}).

Also for the PQ we have not yet studied the consequences when $T$ is a stopping time \cite{Edgar-stopT,Edgar-QM-stopping-time-PRL2019,Neri-stopT}. Often the many advantages of martingale theory come with this concept.
With wider scope, hidden martingales under non-Markovian processes mentioned at the end of Section \ref{sec:model} should be exploited in concrete evolution models beyond the quasi-static protocol.

\paragraph{Note added in proof.} In fact during the protocol of PQ, the quasi-equilibration is not necessary. We were able to 
show this very recently both analytically and numerically.

\acknowledgments 
We deeply thank Guilhem Semerjian for a clarifying comment on the canonical structure.

\bibliographystyle{apsrev4-2.bst}

\end{document}


\title{Supplemental Material to ``Martingale-induced local invariance in progressive quenching''}
	\author{Charles Moslonka}
	\affiliation{Laboratoire Gulliver, UMR CNRS 7083, ESPCI Paris, Université PSL\\
		10 rue Vauquelin, 75005, Paris, France.}
	\email[Corresponding author: ]{charles.moslonka@espci.psl.eu}
	\author{Ken Sekimoto}
	\affiliation{Laboratoire Gulliver, UMR CNRS 7083, ESPCI Paris, Université PSL\\
		10 rue Vauquelin, 75005, Paris, France.}
	\affiliation{Laboratoire Matière et Systèmes Complexes, UMR CNRS 7057, Université de Paris,\\
		10 Rue Alice Domon et Léonie Duquet, 75013, Paris, France }
	\email[Corresponding author: ]{ken.sekimoto@espci.psl.eu}
	
	\date{\today}
	
\maketitle
\onecolumngrid

\setcounter{equation}{0}
\setcounter{figure}{0}
\setcounter{table}{0}
\setcounter{page}{1}
\makeatletter
\renewcommand{\thesection}{S\arabic{section}}
\renewcommand{\theequation}{S\arabic{equation}}
\renewcommand{\thefigure}{S\arabic{figure}}

\section{Hidden martingale for any homogeneous Ising spin dynamics
\label{sec:tower-rule}}
Suppose an Ising system $\{s_k\}_{k=1}^{N_0}$ that evolves under the rule such that each unquenched spins are statistically equivalent (the system is then said ``homogeneous''). For example, when the spins up to $T$-th, $\{s_T,\ldots,s_1\},$ with $0\le T<N_0$ have been quenched, the expectation
$m_{T,M_T}\equiv E[s_{T+1}|M_T,\ldots,M_0]$ should be
 equal to $E[s_{N_0}|M_T,\ldots,M_0].$ 
 Here the conditional expectation $E[s_{N_0}|M_T,\ldots,M_0] (=m_{T,M_T})$ should be defined in the path space such that
the condition, ``$M_T,\ldots,M_0$'', represents the { history} of quenching under a given time protocol, and that the value of $s_{N_0}$, or equivalently that of $s_{T+1},$ should be observed at the right moment when the latter spin is quenched.
The standard martingale theory tells  : {\it If $Z$ and $\{Y_i\}$'s are random variables with finite expectations, the process $\{X_T\}$ defined by $X_T=E[Z| Y_T,\ldots, Y_0]$ is martingale with respect to $\{Y_T,\ldots,Y_0\}$.} In order to prove this, it suffices to use the following property called the {\it tower rule}: $E[E[Z|Y_{T+1},Y_T\ldots,Y_0]|Y_{T},\ldots,Y_0]]= E[Z|Y_{T},\ldots,Y_0].$ In our context, we apply the mapping, $\{Y_T,\ldots,Y_0\}\mapsto \{M_T,\ldots, M_0\}$ (or, equivalently $\{s_T,\ldots, s_1\}$), $X_T\mapsto m_{T,M_T}$ and $Z\mapsto s_{N_0}$ so that
\begin{equation} \label{eq:mdef}
    m_{T,M_{T}}=  E[s_{N_0}|M_{T},\ldots,M_0]
\end{equation}
Then, combining this with the homogeneity mentioned above, we have
\beqa \label{eq:meqtower}
E[m_{T+1,M_{T+1}}|M_{T},\ldots,M_0]  
&=& E[ E[s_{N_0}|M_{T+1},\ldots,M_0] |M_{T},\ldots,M_0]
\cr &=& E[s_{N_0}|M_{T},\ldots,M_0] 
\cr &=& m_{T,M_T}.
\eeqa
Therefore, $m_{T,M_T}$ is martingale process with respect to 
$\{M_T,\ldots, M_0\}$ even if the evolution is not Markovian. 
Note, however, that the results (5) and (6) hold only when $M_{T}$ is Markovian  while (3) or (4) in the main text are valid under non-Markovian processes, $M_{T}.$ Our study in the main text is for the quasi-equilibrium quenching, which is Markovian. Then (\ref{eq:mdef}) reads
$   \meq_{T,M_{T}}=  E[s_{N_0}|M_{T}] .$

\section{Canonical sub-distribution \label{sec:can-sub}}

Let us denote by $\sigma_{0,0}$ all the configurations of the Ising spins, $\{s_1,\ldots,s_{N_0} \},$ and by  $\sigma_{T,M}$ the sub-ensemble of spin configurations under the constraints,
 $s_1=\ldots =s_{\frac{T+M}{2}}=+1$
and  $s_{\frac{T+M}{2}+1}=\ldots =s_{T}=-1.$
Then we define the sub-partition function $Z(\sigma_{T,M})\equiv \sum_{ \{s_1,\ldots,s_{N_0}\}\in \sigma_{T,M}} e^{-H}$ (we have taken the unit of energy such that $\beta=(\kT)^{-1}=1$).
By definition, $Z(\sigma_{0,0})$ is the full partition function.
Now the canonical probability $P^{ (can)}(T,M)$ of observing that $\sum_{i=1}^T s_i=M$ is
$\sigma_{T,M}$ among $\sigma_{T,M}$ reads
\beqa \label{eq:sub-can}
P^{ (can)}(T,M) &=& \binom{T}{\frac{T+M}{2}}
\frac{Z(\sigma_{T,M})}{Z(\sigma_{0,0})}
\cr &=&
\mbox{cst.} \binom{T}{\frac{T+M}{2}}
\intII e^{-\frac{j}{2N_0}(y^2-M^2)}\inRbracket{2 \cosh\inSbracket{\frac{j}{N_0}(y+M)}}^{N_0-T} dy
\cr &=& 
\mbox{cst.} \binom{T}{\frac{T+M}{2}}\sum_{k=0}^{N_0-T}\binom{N_0-T}{k}
e^{\frac{j}{2N_0}(2k-N_0+T+M)^2 }
\eeqa
where the Hubbard-Stratonovich transformation \cite{hubbard1959calculation} has been used to do the sum over the allowed spin configurations and `cst.' is a constant independent of $T$ and $M.$
Especially the result is simple and well known for $T=N_0,$ where
$$P^{ (can)}(N_0,M) = \mbox{cst.} \binom{N_0}{\frac{N_0+M}{2}}  e^{\frac{j}{2N_0}M^2}.$$
We will use the last result in {\bf S3} below.
The summation on the r.h.s. of (\ref{eq:sub-can}) is a sub-partition function the unquenched spins and it is rather surprising that it can be represented as an invariant product in Eq.(6) in the main text.

\section{Bimodality of Distribution under Critical Coupling
\label{sec:largeN0}
}
\newcommand{\jeff}{{j_{\rm eff}}}

We present an argument to assert that the distribution $P(T=N_0,M)$ remains bimodal in the asymptotic limit $N_0\to\infty.$ Even though the split is of order close to $\mathcal(\sqrt{N_0})$ the limit distribution is not Gaussian.

\paragraph{Preparation of $\meq_{T,M}$:}
The basic quantity is the mean equilibrium spin $\meq_{T,M}$ which is defined by $\meq_{T,M}= \frac{\partial}{\partial h}\inRbracket{ \frac{\log Z}{N}},$ where 
$Z$ is the partition function for the $N=N_0-T$ Ising spins with
the pair coupling $\frac{j}{N_0}$ and under the ``molecular field'' $h\equiv \frac{j}{N_0}M.$ Using the Hubbard-Stratonovich transformation it reads
$Z=\log \int e^{N\psi(m)}dm,$ where 
$$ 
\psi(m):= -\frac{\jeff}{2} m^2 +\log [\cosh(h+\jeff m)],
$$
$$ 
\jeff:= \frac{N}{N_0}j =\inRbracket{1-\frac{T}{N_0}}j.
$$
We will use the saddle-point approximation for the integral in $Z,$ 
 which is valid for $T$ such that  $\mathcal{O}(\frac{N}{N_0})=1$ and also $\mathcal{O}(\frac{T}{N_0})=1.$ 
$$
\int e^{N\psi(m)}dm\simeq e^{N\psi(m^*)} \sqrt{\frac{2\pi}{N\psi''(m^*)}},
$$
$$  m^*=\tanh (h+\jeff m^*),
$$
where the second equation defining $m^*$ originates from $\psi'(m^*)=0.$ Using the formulas of $\tanh$ etc.
we can show the formula like,
$\frac{\partial \psi(m^*)}{\partial h}=m^*,$
$\psi''(m^*)=  -\jeff +(\jeff)^2[1-(m^*)^2],$ 
$\frac{\partial m^*}{\partial h} =\frac{1-(m^*)^2}{1-\jeff (1-(m^*)^2)}.$
Combining these, we arrive at a closed equations giving $\meq_{T,M}$:
{\beqa 
\label{eq:meq2}
\meq_{T,M}&=&
m^* -\inv{2N} \frac{\frac{j N}{N_0}  m^* [1-(m^*)^2]}{\inRbracket{1- \frac{jN}{N_0} [1-(m^*)^2]}^2}, 
\qquad \cr 
 && 
m^*= \tanh\inRbracket{\frac{j M}{N_0}+\frac{j N}{N_0}m^*}
\eeqa}
The condition $\mathcal{O}(\frac{T}{N_0})=1$ is also necessary when $j$ is close to the critical value, which is 1 for $N_0\to\infty.$ Since $j$ appears through $\jeff=j N/N_0 ,$ the denominator of the second term on the r.h.s. of the first equation in (\ref{eq:meq2}) 
can become very small for small values of $T,$ invalidating (\ref{eq:meq2}).
In assuming both  $\mathcal{O}(\frac{N}{N_0})=1$ {\it and}
$\mathcal{O}(\frac{T}{N_0})=1,$ we will ignore this term as $\mathcal{O}(N_0^{-1}).$ In short, in (\ref{eq:meq2}) the $\mathcal{O}(N^{-1})$ term  serves only for detecting its validity limit.

\paragraph{Split of the maxima of probability, $M^\circ(T).$}
The extrema $M=M^\circ(T)$ of the probability distribution $P(T,M)$ is found from 
Eq. (14) 
such that $P(T,M^\circ-1)=P(T,M^\circ+1).$
The result reads 
\beq
\label{eq:split-formula}
\meq_{T, M^\circ(T)}=\frac{M^\circ(T)}{T+2}.
\eeq
(We generalize this condition for non-integer $M^\circ$ because $M^\circ\gg 1$ for large $N_0.$)
Apparently $M^\circ(T)=0$ is always the solution by the symmetry reason.
Besides, if $M^\circ >0$ exists, $(-M^\circ)$ does also.
Using (\ref{eq:meq2}) with ignoring the second term on the r.h.s. of the first equation, (\ref{eq:split-formula}) becomes 
we have 
$$\frac{M^\circ(T)}{T+2}=\tanh\inSbracket{j(1+\frac{2}{N_0})  \frac{M^\circ(T)}{T+2} }.$$
It tells that  $M^\circ(T)/(T+2)$ is independent of $T.$ This linearity, $M^\circ(T)\propto (T+2)$ for $\mathcal{O}(\frac{N}{N_0})=1$  and $\mathcal{O}(\frac{T}{N_0})=1,$ is verified by direct calculation of the distributions.
Anticipating that $M^\circ(T)/(T+2)\ll 1$ for $N_0\gg 1$ we can use $\tanh z\simeq z-\inv{3}z^3.$ Especially, when $0<j-1\ll 1$ we have
\begin{equation}\label{eq:Peaks}
    \frac{M^\circ(T)}{T+2}\simeq \sqrt{3\inRbracket{\frac{2}{N_0}+j-1}}
\end{equation}
and the result is consistent, i.e., $M^\circ(T)/(T+2)\ll 1.$

If we use  $j$ at the ``critical value,'' $j_{\rm crit}(N_0)\simeq 1+ \frac{c}{(N_0)^\nu}$ with $c=5.06$ and $\nu =0.933$ according to \cite{PQ-KS-BV-pre2018}, the above approximation expects $M^\circ(T)\simeq \alpha(N_0) \frac{T+2}{N_0}$ with $\alpha(N_0)=\sqrt{3(2+c N_0^{1-\nu})}N_0^{\inv{2}}.$
\begin{figure}
    \centering
    \includegraphics[width=0.6\textwidth]{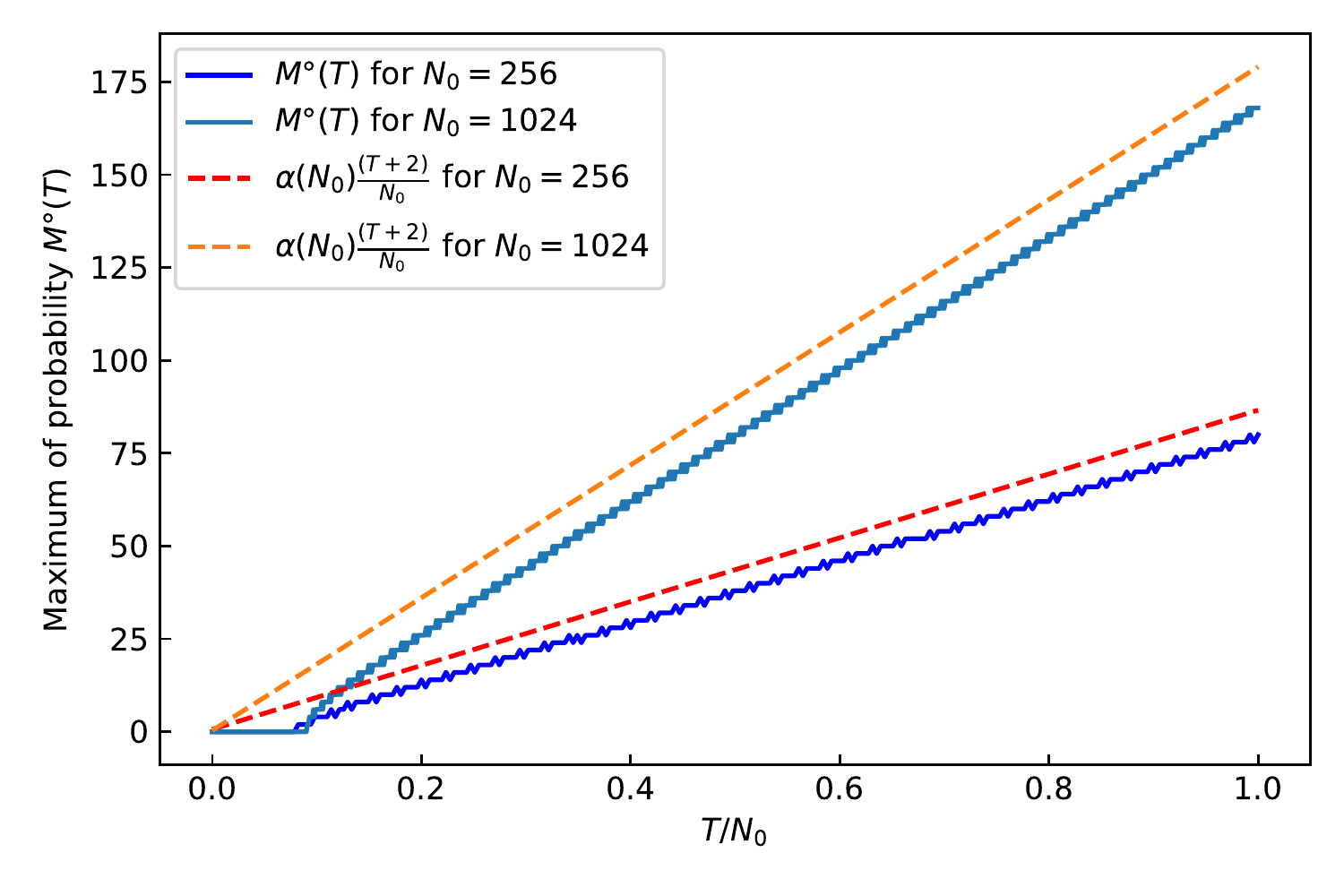}
    \caption{Position of the bimodal peak of $P^{[\infty]}(T,M)$ versus  $\frac{T}{N_0}$ for $N_0=256$ (bottom curve) and $N_0=1024$ (top) obtained numerically ({\it solid curves}).
    The {\it dashed red curves} show the respective asymptotic formula Eq.(\ref{eq:Peaks}).
    }
    \label{fig:peaks}
\end{figure}
In Fig.\ref{fig:peaks} we show the numerical result for $M^\circ(T)$ vs $\frac{T}{N_0}$ for different sizes, $N_0, $ without the saddle-point approximation. What we observed so far is that, once the bimodality appears at some stage of progressive quenching, $T=T_0 (<N_0),$ it remains for any $T$ with $T_0\le T \le N_0.$ Admitting this as a fact, we conclude that 
the bimodality of $P(T=N_0,M)$ remains and we expect $M^\circ(N_0)\simeq
\sqrt{3c} N_0^{\frac{2-\nu}{2}}.$
This claim is consistent with the claim that $\vec{P}^{ (PQ)}(T)$ 
that started from the condition $\vec{P}^{ (PQ)}(0)=\{1\}$ 
is the canonical weight for the sub-distribution $\vec{P}^{ (can)}(T).$ 
In fact for $T=N_0$ such canonical weight (\ref{eq:sub-can}) is analytically tractable and read for $N_0\gg 1$ as follows:
$$ 
{P}^{ (can)}(N_0,M)= \mbox{cst.} \binom{N_0}{\frac{N_0+M}{2}}  e^{\frac{j}{2N_0}M^2}
\simeq \sqrt{\frac{2}{\pi(1-\mu^2)}}e^{-N_0 \phi(\mu;j)}
$$
with $\mu=\frac{M}{N_0}$ and $\phi(\mu;j)$ being the rate function of the large deviation principle given by
$$ \phi(\mu;j)=\frac{j}{2}\mu^2 -\frac{1+\mu}{2}\log(1+\mu)-\frac{1-\mu}{2}\log(1-\mu)
\simeq \frac{1-j}{2}\mu^2+\frac{\mu^4}{12}.$$ 
This also gives $\frac{M^\circ}{N_0}\simeq \sqrt{3(j-1)}$ in the limit $N_0\to \infty,$ being consistent with (\ref{eq:Peaks}). Because $1-j_{\rm crit}(N_0)\to 0$ under $N_0\to \infty,$ the PQ makes the distribution approach to the critical one $P^{\rm (can)}(\infty, M)$ as limit of bimodal distribution. 

As for small values of $T,$ the full numerical results show the unimodal-bimodal transition with $T$, see Fig.\ref{fig:peaks} (thick curves).

\section{ Calculation of transfer matrices under \texorpdfstring{$\bm{K}$}{K},
\texorpdfstring{$\bm{S}$}{S}, \texorpdfstring{$\bm{KS}$}{KS} and \texorpdfstring{$\bm{SK}$}{SK}}
\label{sec:app-OP}
In this section, we derive the transfer matrix elements for the probability vector under the operation of progressive quench $\bm{K},$
unquenching of randomly selected spin $\bm{S},$ as well as their 
combinations $\bm{KS}$  and $\bm{SK}.$

We will use the symbol $\delta(\, )$ for the Kronecker's delta, i.e., $\delta(n)$ with $n\in \Z$ takes the value 1 for $n=0$ and 0 otherwise. We also write
the conditional expectation using the symbol $E$ such as 
$E[X | Y]$ for the expectation of $X$ given the knowledge of $Y.$  When $Y$ is a random variable, $E[X|Y]$ does also. 
The component $P(T,M)$ of the probability vector $\vec{P}(T)$ reads
$$P(T,M)=E[\delta(M-\hat{M}_T)].$$

We will abuse the operators $\bm{K}$ and $\bm{S}$ to act both on the quenched magnetization $\hat{M}_T$ when $T$ spins are fixed and also on the 
probability vector $\vec{P}(T),$ i.e., on the ensemble of systems having 
different $M_T$ according the given weights.
When  $\bm{L}$ stands for the operators, $\bm{K},$ $(\bm{KS}),$ etc.,
\beqa \label{eq:rule}
E[\delta (M-\bm{L}\hat{M}_T ) | \hat{M}_T] 
&=&
\sum_{k}\delta(M-(\hat{M}_T+k))  a_{T,k}(\hat{M}_T)
\cr
&=& 
\sum_{k}\delta((M-k)-\hat{M}_T)  a_{T,k}(M-k),
\cr &&
\eeqa
where $a_{T,k}(\,)$ are the weights, can be translated into
the usual representation in terms of the transfer matrix elements 
as 
\beq
\bm{L}P(T,M)
=
\sum_{k}P(T,M-k)  a_{T,k}(M-k),
\eeq

\paragraph{Operation of $\bm{K}$:} 
As described above $\bm{K}\hat{M}_T$ means the quenched magnetization after a unquenched spin out of $N_0-T$ ones has been quenched. The system then has $T+1$ quenched spin. The conditional distribution of the 
resulting magnetization reads,
\beq \label{eq:dKapp}
    \begin{aligned}
		E[\delta (M-\bm{K}\hat{M}_{T}) | \hat{M}_T]
        &= \delta(M-(\hat{M}_T+1))\frac{1+\meq_{T,\hat{M}_T}}{2}\\
        &+\delta(M-(\hat{M}_T-1))\frac{1-\meq_{T,\hat{M}_T}}{2}\\
        &=\delta(M-1-\hat{M}_T)\frac{1+\meq_{T,M-1}}{2}\\
        &+\delta(M+1-\hat{M}_T)\frac{1-\meq_{T,M+1}}{2}
    \end{aligned}
\eeq
For the later convenience we rewrite (\ref{eq:dKapp}) with $T\to T-1$.
\beq \label{eq:dK2app}
    \begin{aligned}
        E[\delta (M-\bm{K}\hat{M}_{T-1}) | &\hat{M}_{T-1}]=\\
        \delta(M-1-&\hat{M}_{T-1})\frac{1+\meq_{T-1,M-1}}{2}\\
        +\delta(M+1-&\hat{M}_{T-1})\frac{1-\meq_{T-1,M+1}}{2}
    \end{aligned}
\eeq
or, using the general relationship (\ref{eq:rule})
we find $\bm{K}\vec{P}(T-1)$ as the probability vector in the $T$-sector with the component, 
\beq \label{eq:dK2app0}
    \begin{aligned}
        (\bm{K}\vec{P}(T-1))_M&=P(T-1,M-1)\frac{1+\meq_{T-1,M-1}}{2}\\
        &+P(T-1,M+1)\frac{1-\meq_{T-1,M+1}}{2}
    \end{aligned}
\eeq

\paragraph{Operation of $\bm{S}$:} 
We denote by $\bm{S}\hat{M}_T$ the quenched magnetization after 
a quenched spin out of $T$ ones has been unquenched. The system  has
$T-1$ quenched spins and $N_0-T{+1}$ unquenched spins.
The conditional distribution of the resulting magnetization reads:
\beq \label{eq:dSapp}
    \begin{aligned}
        E[\delta(M-\bm{S}\hat{M}_T)|\hat{M}_T]&=
        \delta(M-(\hat{M}_T-1))\frac{1+\frac{\hat{M}_T}{T}}{2}\\
        &+\delta(M-(\hat{M}_T+1))\frac{1-\frac{\hat{M}_T}{T}}{2}\\
        &=\delta(M+1-\hat{M}_T)\frac{1+\frac{M+1}{T}}{2}\\
        &+\delta(M-1-\hat{M}_T)\frac{1-\frac{M-1}{T}}{2}.
    \end{aligned}
\eeq

\paragraph{Operation of $\bm{KS}$:} 
$\bm{K}(\bm{S}\hat{M}_{T})$ means to unquench randomly a spin among $T$ quenched ones then quench randomly a spin among $N_0-(T-1)$ thermalized spins. 
Replacing in (\ref{eq:dK2app})  $\hat{M}_{T-1}$ by $\bm{S}\hat{M}_T,$
where $\bm{S}\hat{M}_T$ is given in (\ref{eq:dSapp}), the result reads :
\beq
\begin{aligned} 
    E[\delta (M&-\bm{K}(\bm{S}\hat{M}_{T})\,) | \hat{M}_{T}]\\ 
    &=\delta(M-\hat{M}_{T}) \frac{1+\frac{\hat{M}_{T}}{T}}{2}
    \frac{1+\meq_{T-1,\hat{M}_{T} -1}}{2}\\
    &+ \delta(M-2-\hat{M}_{T}) \frac{1-\frac{\hat{M}_{T}}{T}}{2}\frac{1+\meq_{T-1,\hat{M}_{T}+1}}{2}\\
    &+\delta(M+2-\hat{M}_{T}) \frac{1+\frac{\hat{M}_{T}}{T}}{2}\frac{1-\meq_{T-1,\hat{M}_{T}-1}}{2} \\
    &+\delta(M-\hat{M}_{T}) \frac{1-\frac{\hat{M}_{T}}{T}}{2}\frac{1-\meq_{T-1,\hat{M}_{T}+1}}{2}
\end{aligned}
\eeq
By taking the expectation over $\hat{M}_T$, i.e. the weighted summation $\sum_{M_T=-T}^T P(M_T,T),$ we have the evolution of $\vec{P}$ after a single cycle of operation, $\bm{KS}.$ 
The fixed point equation (12) 
 in the main text is obtained by requiring $\bm{(KS)}\vec{P}(T)=\vec{P}(T).$
The rewriting this into the form of (12) 
 is very close to 
the transformation from (3) 
 to (4). 
 The close relationship between 
the martingale and the harmonic function has long been known \cite{harmonic-Doob}.

\paragraph{Operation of $\bm{SK}$:} 
$\bm{S}(\bm{K}\hat{M}_{T})$ means to quench randomly a spin among the $N_0-T$ thermalized ones then unquenching randomly a spin among $T+1$ quenched ones. 
In the manner similar to the case of operating $\bm{KS},$ the result reads

\beq
    \begin{aligned}
        E[\delta (M-\bm{S}(\bm{K}\hat{M}_{T})\,) | \hat{M}_{T}]
        &=
        \inSbracket{
        \delta(M-M_T -2 ) \left( \frac{1- \frac{M_T+1}{T+1}}{2}\right)+\delta(M-M_T) 
        \left( \frac{1+ \frac{M_T+1}{T+1}}{2}\right)
        }
        \frac{1+\meq_{T,M_T}}{2}\\
        &+ 
        \inSbracket{
        \delta(M-M_T)
        \left(\frac{1- \frac{M_T-1}{T+1}}{2} \right) +\delta(M-M_T +2 )
        \left(\frac{1+ \frac{M_T-1}{T+1}}{2}\right)
        }
        \frac{1-\meq_{T,M_T}}{2}
    \end{aligned}
\eeq

By taking the expectation over $\hat{M}_T$, i.e. the weighted summation $\sum_{M_T=-T}^T P(M_T,T),$ we have the evolution of $\vec{Q}$ upon 
after a single cycle of operation, $\bm{SK}.$ 
The fixed point equation (13) 
in the main text is obtained by requiring $\bm{(SK)}\vec{Q}(T)=\vec{Q}(T).$


\bibliographystyle{apsrev4-2.bst}